# A light-weight and high thermal performance graphene heat pipe

Ya Liu,[1,2] Shujing Chen,[3] Yifeng Fu,[1] Nan wang,[4] Davide Mencarelli,[5] Luca Pierantoni,[5] Hongbin Lu,[2*] Johan Liu [1,3*]

[1]Electronics Materials and Systems Laboratory, Department of Microtechnology and Nanoscience, Chalmers University of Technology, Kemivägen 9, SE 412 96 Gothenburg, Sweden

[2] State Key Laboratory of Molecular Engineering of Polymers, Department of Macromolecular Science, Collaborative Innovation Center of Polymers and Polymer Composites, Fudan University, 2005 Songhu Road, Shanghai 200433, China

[3]SMIT Center, School of Mechanical Engineering and Automation, Shanghai University, No 20, Chengzhong Road, Shanghai, Box 808, 201800, China

[4]SHT Smart High Tech AB, Kemivägen 6, SE 412 58 Gothenburg, Sweden

[5]Università Politecnica delle Marche (UNIVPM), Via Brecce Bianche, Ancona, Italy

Email: johanliu@shu.edu.cn; hongbinlu@fudan.edu.cn; johan.liu@chalmers.se

**Abstract**

Heat pipe is one of the most efficient tools for heat dissipation in electronics and power systems. Currently, heat pipes are usually made of copper, aluminum or stainless steel. Due to their relatively high density and limited heat transmission capacity, heat pipes are facing urgent challenges in power electronics and power modules. In this paper, we report a new class of graphene enhanced heat pipes that can cope with these issues. The graphene enhanced heat pipes are made of high thermal conductivity graphene assembled film and graphene laminated copper films with nanostructure enhanced inner surfaces. The study shows that the dramatically improved heat dissipation capacity, 6100 W m$^{-2}$ K$^{-1}$ g$^{-1}$, about 3 times higher than that of copper based commercial heat pipes can be achieved. This paves the way for using graphene enhanced heat pipes in light-weight and large capacity cooling applications, as required in many systems such as avionics, automotive electronics, laptop computers, handsets and space electronics.

Nowadays, the power density in integrated circuits (ICs) and hotspots can reach over 1000 W m$^{-2}$.[1] Efficient heat dissipation is essential for keeping the performance and extending the lifetime of electronics and power systems. Heat pipe, a two-phase flow heat transfer device, is emerged as one of the most important methods because of its high efficiency and unique ability to transfer heat over large distance with minimal losses.[2-4] Today, heat pipes are mainly made of metals with high thermal conductivity and good mechanical strength, such as copper[5], aluminum[5,6] or stainless steel[7,8] However, with the development of terrestrial and portable electronics, these metals could no longer be the best choice because light-weight has become the first priority.[5] Several attempts have been made to lighten heat pipes by using light-metal and alloys, epoxy-impregnated carbon fiber with thin aluminum shells and metal/matrix composites.[5] However, light-metal suffers from issues such as corrosion and low thermal conductivity [9], while the incorporation of epoxy or polymer matrix sharply increases thermal transfer resistance.[10] Therefore, it remains an urgent challenge to develop a light-weight and high thermal performance heat pipe.

Compared to metal materials, graphene shows overwhelming advantages such as high thermal conductivity, light-weight, good stability and superior in-plane thermal conductivity (5300 W m$^{-1}$ K$^{-1}$ at room temperature),[11] far better than the thermal conductivity of copper (402 W m$^{-1}$ K$^{-1}$) and aluminum (237 W m$^{-1}$ K$^{-1}$).[12] Compared to single layer graphene film, graphene assemble films (GAF) reaches a good compromise between thermal conductivity, scalable preparation and applicable mechanical strength;[13,14] for instance, micron-scale GAFs have exhibited thermal conductivities up to 2000-3200 W m$^{-1}$ K$^{-1}$ and tensile strength of 78 ± 6 MPa .[13] Such thermal conductivity is 4-10 times higher than that of copper and aluminum but lighter in weight because of the lower density ( < 2.2 g/cm$^3$).[15] More importantly, unlike metal-based heat pipes, graphene assemble films have little corrosion risks even under acid, alkali and moisture exposure.[16,17]

In this work, we demonstrate a strategy to constitute a new class of light-weight and highly thermal conductive heat pipe based on GAF.[13] The heat pipe exhibits a thermal transfer coefficient up to 16085 W m$^{-2}$ K$^{-1}$ under an input power of 10 W, corresponding to a cooling capability of 6100 W m$^{-2}$ K$^{-1}$ g$^{-1}$, which is about three times better that that of copper based heat pipe with the same geometry (2053 W m$^{-2}$ K$^{-1}$ g$^{-1}$). Simulation results show that the graphene film contributes over 30% to the total heat dissipation ability of the heat pipe due to its outstanding thermal conductivities. Heat transfer modelling suggests that increasing thermal conductivity of container can significantly improve its heat dissipation capacity even though heat is primarily taken away by phase change in heat pipe.

**Graphene heat pipe design and fabrication.**

Our strategy for fabricating graphene heat pipe involves three key components, including container, wicker and working fluid (Fig.1a, b). First, we investigated the microstructure of the GAF by small-angle X-ray scattering (SAXS). It implies that tailing orientation of graphene flakes contributes to achieve high thermal conductivity GAF (Supplementary Fig. 1). Such orientation behaviors provide evidence for the kinetic transport theory of graphene,[19] and thermal conductivity of graphene can be expressed as

$$K = \left(\frac{1}{2}\right) C v \Lambda \quad (1)$$

where $\Lambda$ is the phonon mean free path, C is specific capacity, $v$ is phonon group velocity. [19] The high orientation of graphene flakes enhances the possibility to achieve a higher phonon mean free path for improving its thermal conductivity. Based on such a hypothesis, following the previous route,[13] we prepared a 25 µm thick graphene assembled film with in-plane thermal conductivity as high as 1870 W m$^{-1}$ K$^{-1}$ by tailoring the structure of graphene flakes[13] (Supplementary Fig. 2).

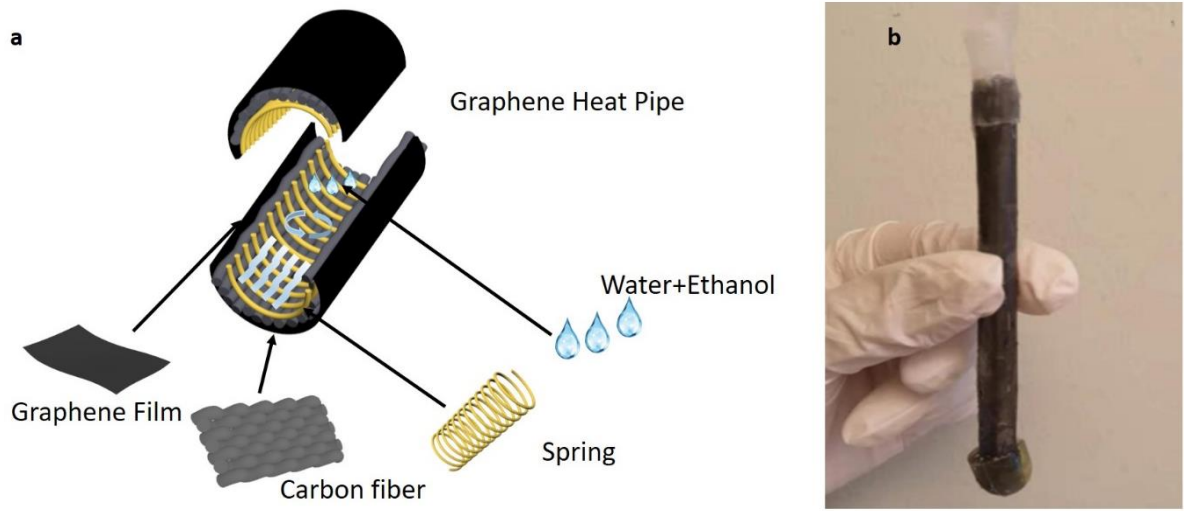

**Fig. 1** Design, image and principle of operation of the graphene heat pipe. **a** Schematic drawing of the graphene heat pipe. **b** Picture of a real graphene heat pipe.

Then, we exploit three porous structures, including GAF, carbon fiber mesh (CFM) and carbon fiber brunch (CFB) to optimize the wick structure. Among these three wickers, the heat pipe with CFB wicker shows the highest effective thermal transfer coefficient and the lowest temperature on heat pipe (Supplementary Fig. 3). Capillary pressures of three wickers were calculated to explain the relationship between wicker structure and heat transfer performance. When the heat pipe is operated in steady state, the capillary pressure of wicker is expressed as

$$\Delta p = 2\delta\cos\theta/r \qquad (2)$$

where $\delta$ is the surface tension of working fluid (23.82 mN m$^{-1}$),[20] $\theta$ is the wetting angle, r is the effective capillary radius or pores of the wicker.[4]

For the three wicker structures, r is calculated as

$$R_{GAF} = (w+d_w)/2 \qquad (3)$$

$$R_{CFM} = (w+d_w)/2 \qquad (4)$$

$$R_{CFB} = w \qquad (5)$$

where w is the wire spacing, $d_w$ is the wire diameter.[4,21]

Microscopic structures (Supplementary Fig. 3e-j) and wetting angles (Supplementary Fig. 4) of three wickers were measured to obtain $\theta$ and r. Therefore, the capillary pressure of three wickers are calculated as 4.76, 6.35, 15.88 KPa. IR camera provided evidence of capillary pressure by visualizing the movement of the working fluid (Supplementary Fig. 5 - 7). In corporation with thermal transfer capability of heat pipes made from the three wickers, it demonstrates that the capillary pressure of CFB has largest contribution to thermal transfer efficiency. Besides, we studied the composition and filling amount of working fluid (Supplementary Fig. 3k, l). Accordingly, 30 vol.% of water/ethanol (1:4) solution is the optimized parameter for the graphene heat pipe.

**Heat dissipation performance of the graphene heat pipe.**

With these optimized parameters, graphene heat pipe with various length (90, 130, 150 mm) were fabricated to investigate the relationship between length and thermal transfer efficiency. When the length of the 6 mm outer-diameter heat pipe decreases from 150 to 90 mm, temperature on the evaporator decreases from 53 to 41 °C (Fig. 2a - c). Meanwhile, the effective thermal transfer coefficient increases with shortened heat pipe length and reaches 16085 w m$^{-2}$ K$^{-1}$ with an input power of 10 W (Fig. 2d). It is clear that the heater temperature decreases with shortened pipe length (Fig. 2e), consistent with thermal transfer coefficient trends. With a 10 W input power, temperature of the heater containing a 90 mm graphene heat pipe goes to 41 °C after 10 mins, far lower than that (64 °C) of the independent heater without graphene pipe. These provide evidence that shorter heat pipe carry more power than longer pipe since capillary limit is an inverse function of the length.[20]

To mimic a real application, the working direction of graphene heat pipe was set to be 45° and 0° (Supplementary Fig.8). For two cases (45° and 0°), the assistance from gravity to convey working fluid reduces, so that the vertically orientated (90°) heat pipe shows the lowest temperature and best thermal transfer efficiency (Fig. 2f). Nevertheless, the heat pipe still achieves thermal transfer coefficients of 11524 and 9084 w m$^{-2}$ K$^{-1}$, respectively, when declining (45°) and horizontally orientated (0°). Such heat transfer performance can meet the cooling demands for most of the portable mobile devices.[22]

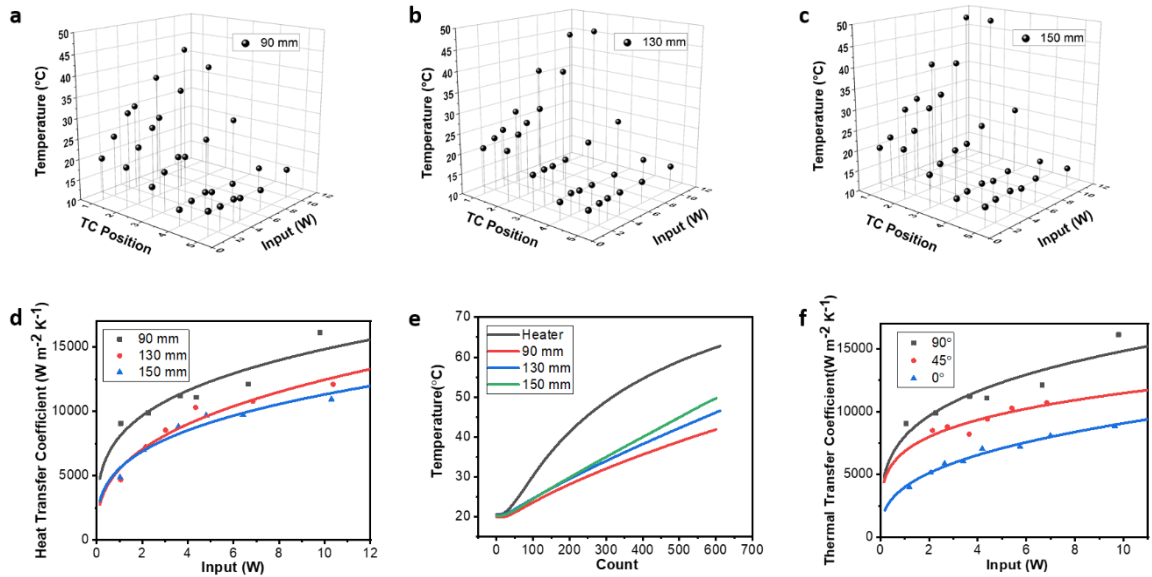

**Fig. 2** Heat dissipation performance of a 6 mm graphene heat pipe. **a, b, c** Temperature distribution along heat pipe with length as 90, 130 and 150 mm; **d** Thermal transfer coefficient of graphene heat pipe with length as 90, 130 and 150 mm; **e** Temperature distribution on independent heater and heater with 90, 130 and 150 mm graphene heat pipe under 10 W input; **f** Thermal transfer coefficient of graphene heat pipe with working direction as 90°, 45° and 0°.

We also fabricated graphene/copper (G/C) composite heat pipe with the same diameter (outer diameter 6mm, length 150 mm), while a commercial copper heat pipe (Spread Fast AB, SF-10-150-S) was characterized as reference. G/C heat pipe shows similar structure with graphene heat pipe (Fig. 3a, c). The G/C composite film was prepared by electroplating (Supplementary Fig. 9) with thickness of both graphene and copper is about 25 µm (Fig. 3b). Thermal conductivity of such G/C film is 477 W m$^{-1}$ K$^{-1}$.

Sintered copper particle works as a wick structure in the commercial copper heat pipe (Fig. 3d, e). Temperature distributions on the graphene/copper composite heat pipe and commercial copper-based heat pipe are shown in Supplementary Fig. 10, 11. Compared to the commercial copper heat pipe, the graphene and graphene/copper composite heat pipes show a significant advantage on thermal transfer coefficient per weight (Fig. 3f). The specific thermal transfer coefficient of graphene heat pipe is improved by 3 times compared to that of the copper heat pipe. Such a high specific thermal transfer coefficient makes graphene heat pipe an ideal candidate for thermal management on lightweight applications such as in spacecraft, avionics, automotive and consumer systems where performance vs weight is of great concern.

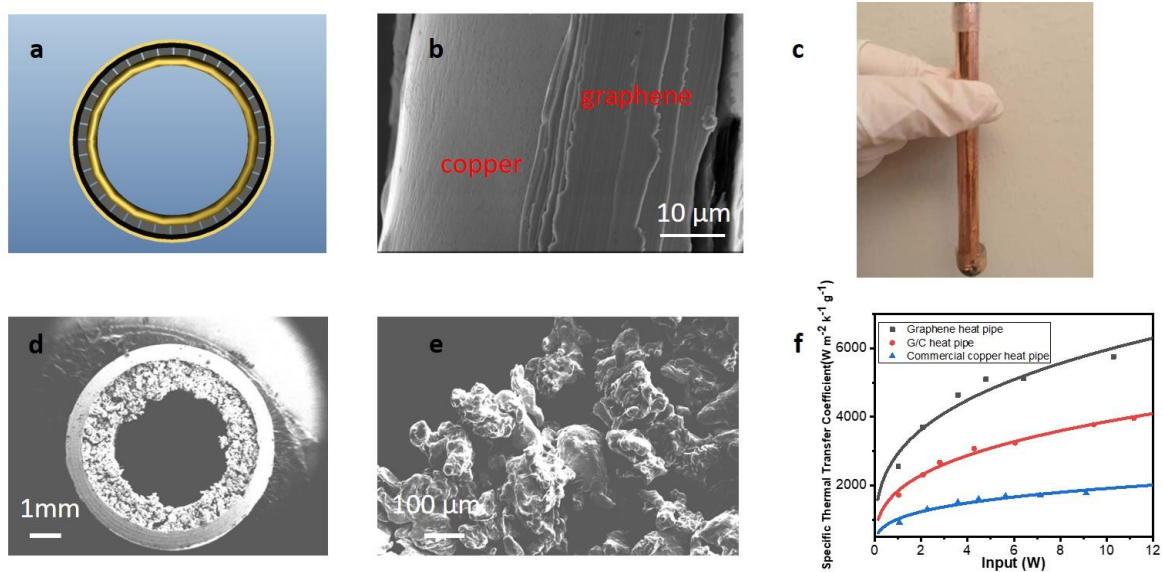

**Fig. 3** Graphene copper composite heat pipe and comparison with a commercial copper heat pipe. **a** Cross section image of G/C heat pipe; **b** SEM images of G/C composite film; **c** image of graphene copper heat pipe; **d** Cross-section SEM image of commercial copper heat pipe; **e** SEM images of wick structure of commercial copper heat pipe; **f** Specific thermal transfer coefficient of graphene, G/C and commercial copper based SF-10-150-S heat pipe.

**Heat transfer modelling for graphene heat pipe.**

The contributions from the container's thermal conductivity and phase change process were assessed numerically by COMSOL Multiphysics solver. Models of 6 mm outer diameter graphene heat pipes with length of 90, 130 and 150 mm have been designed by COMSOL and set for simulation (Fig.4a). In particular, we set here thermal conductivity of container as 400, 900 and 1400 W m$^{-1}$ K$^{-1}$ to figure out the contribution from the container in a graphene heat pipe. Similar to the experimental results, it was found that the shorter heat pipe, the lower temperature is on the heat pipe. Additionally, with the same heat pipe diameter and length, by improving the thermal conductivity of graphene film from 400 to 1400, the temperature on the heater can be cooled from 85.6 to 59.2 °C under 20 W input (Fig. 4c - e). This implies that improving thermal conductivity of the container would effectively promote heat dissipation of the heat pipe. To quantify such an improvement, the contribution factor ($ɣ_c$) of the container has been calculated (Fig. 4b inset). Specifically, $ɣ_c$ of the graphene film exhibits a negatively linear relationship with the length of the graphene heat pipe. While reducing the pipe length, the heat conducted by the container increases. Hence, it explains why the shorter graphene heat pipe, the better its heat transfer performance. Meanwhile, it is observed that the higher thermal conductivity, the higher $ɣ_c$ from the container. As a result, when thermal conductivity of a graphene film reaches 1400 W m$^{-1}$ K$^{-1}$, $ɣ_c$ can goes up to 0.31. This explains why highly thermally conductive graphene film can help achieve high-efficiency heat pipe. Based on these, we build the heat transfer model for the graphene heat pipe.

The total transferred heat is expressed as

$$Q=Q_c+Q_{pc} \qquad (6)$$

Here, $Q_c$, $Q_{pc}$ are heat transferred from conduction and phase change, respectively. According to the definition of thermal conductivity and Riedel equation,

$$Q_c=AΔTλ/L \qquad (7)$$

$$Q_{pc}= 1.093mRT_c\ln(P_c-1)/18(0.93-T_b) \qquad (8)$$

where A is cross-section area of a container, ΔT is temperature difference between evaporator and condenser section, λ represents thermal conductivity of the graphene film, L is the length of the heat pipe, m is the weight of working fluid, R is gas constant, $T_c$ is tested temperature, $P_c$ is critical pressure, $T_b$ is reduced temperature at the normal boiling point. [23,24]

For a certain case, container's contribution factor is

$$ɣ_c = AΔTλ(0.93 − T_b)/[(0.93T_bAΔTλL) + 1.093mRT_cLln(P_c − 1)] \quad (9)$$

While inset heat parameters, container's contribution factor is rewritten as

$$ɣ_c = 1.44λ(0.93 − T_b)/[(0.0168λ + 191T_c\ln(P_c − 1)] \qquad (10)$$

Because of $191T_c ln(P_c - 1) \gg 0.0168\lambda$, container's contribution factor is simplified as

$$ɣ_c = 1.44\lambda(0.93 - T_b)/191T_c ln(P_c - 1) \tag{11}$$

$T_b$ is known. Therefore, it suggests that increase thermal conductivity of container can effectively improve thermal transfer efficiency by improving $ɣ_c$.

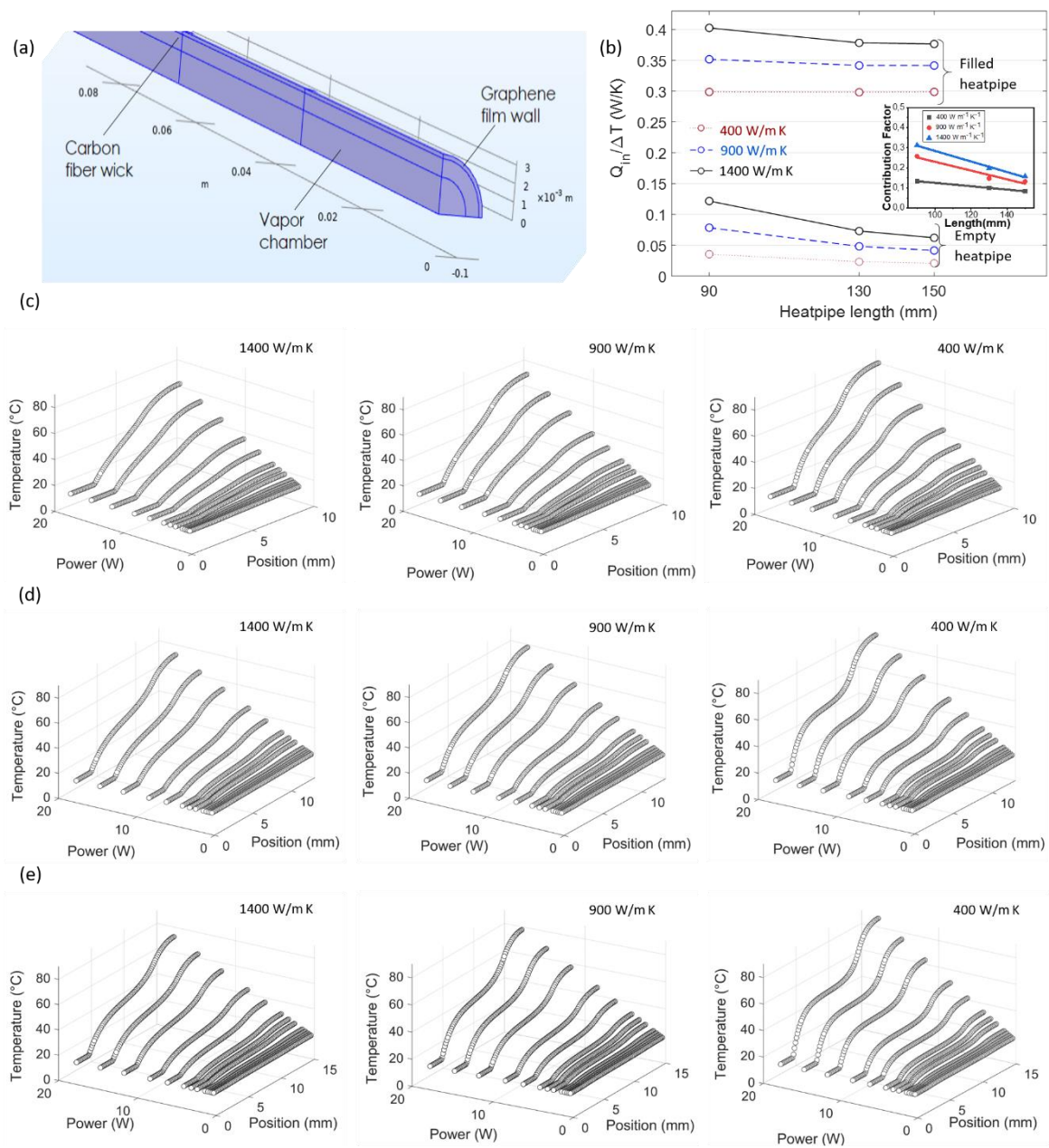

**Fig. 4** Modeling of graphene heat pipe. **a** COMSOL simulation image of graphene heat pipe; **b** Thermal transfer performance of phase change and GAF; **c** Simulation results of 90 mm graphene heat pipe with thermal conductivity as 1400, 900 and 400 W m$^{-1}$ K$^{-1}$; **d** Simulation results of 130 mm graphene heat pipe with thermal

conductivity as 1400, 900 and 400 W m$^{-1}$ K$^{-1}$; **e** Simulation results of 150 mm graphene heat pipe with thermal conductivity as 1400, 900 and 400 W m$^{-1}$ K$^{-1}$.

**Conclusions**

The heat dissipation ability demonstrated by the graphene heat pipe beats the modern well-designed copper heat pipe. The thermal transfer coefficient of the graphene heat pipe is 16085 W m$^{-2}$ K$^{-1}$ under an input power of 10 W. Taking both thermal transfer coefficient and weight into account, the graphene heat pipe significantly outperformed the commercial copper heat pipe. Moreover, copper can be electroplated on the GAF to realize graphene-copper composite heat pipe. Such composite heat pipes are able to well fulfill the requirement for high strength and efficient graphene heat pipes.

The fabrication of graphene heat pipe is scalable and could be used to make dense heat pipe arrays. The preparation of the GAF and the associated electroplating technique are already well-developed for high-yield commercial manufacturing. Therefore, only an integrated sealing setup with a vacuum pump was needed to realize the fabrication of the graphene heat pipes at a large scale. Fortunately, such vacuum equipment is widely used in copper heat pipe industry now. Furthermore, although we measured the graphene heat pipes with the assistance of cooling water in this work, the condenser section can be substituted by a heat sink or fan to make the cooling even more efficient when applied in a real case. Notably, the graphene heat pipe could bear tough situations where metal could be corroded, such as acidic and oxidative environments. Taking all these factors into consideration, the graphene heat pipe shows great advantages and potential in cooling of a variety of electronics and power systems.

To conclude, we for the first time demonstrated the cooling application of graphene (composite) heat pipe. With optimized wick structure, graphene heat pipe exhibits 3 times higher efficiency than the well-designed commercial copper based heat pipe. It is believed that it opens the possibility of boosting the application of graphene for thermal management, especially in the situation of light-weight and corrosion resistant cooling of devices and systems such as in power systems, avionics, space electronics, automotive electronics, and laptop computers as well as handsets.

## Methods

**Preparation of graphene heat pipe.** At first, a high thermal conductivity GAF was prepared from graphene oxide using the reported approach. [13] To remove coated polymers, carbon fiber was soaked into acetone for 2 days. After drying, the carbon fiber was transferred into a plasma chamber at 50 W for 1 min to oxidize its surface. Then carbon fibers were attached to GAF to act as wick structure by a waterproof adhesive (Plexus MA300). A copper spiral was put inside the graphene pipe to hold the pipe during vacuuming. After that, one end of the pipe was sealed by two-component epoxy, while a V-shape plastic pipe was attached on the other end of the graphene pipe. Then the V-shape part was connected to a vacuum system to exhaust the inside air ( Supplementary Fig. 12). After that, valve 1 was closed and valve 2 was turned on to fill in a certain amount of water-ethanol solution. Finally, a heat gun was used to seal the V-shape plastic pipe. The heat pipe was soaked in water to check its tightness during vacuum. All wicker materials were treated by $O_2$ plasma.

**Preparation of graphene/copper heat pipe.** Bright copper plating solution and copper plate were purchased from Tifoo-Electroplating & Surface Technology, Germany. Before electroplating, a 2 nm Tungsten (W) was deposited on GAF. In a working situation, the copper plate was connected to the positive of the DC supplier while GAF works as negative electrode. The current for electroplating is 0.5 A. While finished electroplating, graphene/copper composite film was sintered in a tube furnace under 1000 °C for 60 mins. After that, the graphene copper heat pipe was prepared with the same procedures of the as-prepared graphene heat pipe.

**Measurement of heat pipe performance.** As shown in Supplementary Fig. 13, five thermal couples were attached on the position of 1, 2, 3, 4 and 5. In a typical experiment, an enclosure heating element (70W, RS component Co. Ltd, Sweden) was used as a heater. The heater was connected to a DC power supply (Agilent E3612A). Heat pipe was embedded in the middle of the heater with the assistance of an aluminum box. Thermal grease (Loctite TG100, 3 W m$^{-1}$ K$^{-1}$, China) was used to fill in the gap between heat pipe and aluminum box. As to the condenser section, a copper blocker was tightly connected with the cooling water system. Then the other end of the heat pipe was embedded inside of the copper block.

SEM images were obtained on a Zeiss Supra 60 VP with acceleration voltage 15 kV.

**Simulation.** COMSOL Multiphysics was used to simulate the performance of graphene heat pipe, assuming the same geometrical and physical parameters of the experiments. Heat transfer in wicker is simplified as conduction with an effective heat transfer coefficient. In the simulation, the effective thermal conductivity of carbon fiber wickers was calculated by the following expression:

$$k_{eff} = \frac{k_f(k_f+k_s-(1-\varphi)(k_f-k_s))}{k_f+k_s+(1-\varphi)(k_f-k_s)} \quad (12)$$

where, $k_{eff}$ is the effective conductivity; $k_f$ is conductivity of fluid material (Water/Ethanol mixture); $k_s$ is the thermal conductivity of solid carbon; $\varphi$ is the effective porosity of CFB. Vapor density was assumed as ideal gas as below:

$$\rho = \frac{p}{R_s T} \quad (13)$$

where $\rho$ is vapor density; $R_s$ is gas constant; T is temperature. Indicating with $\lambda$ the latent heat, with $p_{ref}$ the room pressure, and with $T_{ref}$ the evaporation temperature of the mixture, the inlet flow at evaporator side of wick/vapor interface and the outlet flow at condenser side are self-consistent with the saturation pressure given by:

$$p = p_{sat}(T) = p_{ref} \cdot exp\left(\frac{\lambda}{R_s}\left(\frac{1}{T_{ref}} - \frac{1}{T}\right)\right) \quad (14)$$

# Acknowledgements

The authors acknowledge the financial support from the Swedish Board for Strategic Research (SSF) with the contract No: SE13-0061 and GMT14-0045, from the Swedish Board for Innovation (Vinnova) under the Siografen program, from Formas with the contract No: FR-2017/0009 and from STINT for the double degree PhD collaboration program with the contract No: DD2016-6502, from the Swedish National Science Foundation with the contract No: 621-2007-4660 as well as from the Production Area of Advance at Chalmers University of Technology, Sweden. S.C. and J.L. also acknowledge the financial support by the Key R&D Development Program from the Ministry of Science and Technology of China with contract No: 2017YFB0406000 as well as from the National Natural Science Foundation of China (No: 51872182). H.L thanks for the financial support from Shanghai International Collaboration research project (No: 19520713900) and State Key Laboratory of Molecular Engineering of Polymers at Fudan University.